\documentclass[aps,prl,twocolumn,groupedaddress]{revtex4-2}

\usepackage{graphicx}
\usepackage[inkscapelatex=false]{svg}
\usepackage{dcolumn}
\usepackage{bm}
\usepackage[hidelinks]{hyperref}
\usepackage[mathlines]{lineno}
\usepackage{comment}
\usepackage{amsmath}

\usepackage[capitalize, nameinlink, noabbrev]{cleveref}
\crefname{figure}{Fig.}{Figs.}
\crefname{equation}{Eq.}{Eqs.}
\crefname{subfigure}{Fig.}{Figs.}
\labelformat{subfigure}{#1} 

\renewcommand{\d}{\,\mathrm{d}}

\begin{document}

\title{Contacts to Low-Dimensional Semiconductors: Physical Theory and Analytical Model}

\author{Jimmy Qin}
\email{jimmyqin@stanford.edu}
\author{H.-S. Philip Wong}%
 
\affiliation{%
Department of Electrical Engineering, Stanford University, Stanford, CA 94305, USA
}%

\begin{abstract}
    Metal contacts to low-dimensional semiconductors are critical for nanoelectronics, yet a general physical description has remained elusive. We present an analytical model for metal-induced gap states (MIGS), revealing a universal scaling law governed by semiconductor dimensionality. Linking MIGS to transport observables, we provide a unified formulation of Schottky barrier height, transfer length, and contact resistance. Our model explains recent experiments on carbon nanotubes and 2D materials, clarifying the fundamental criteria for achieving scalable, low-resistance contacts.
\end{abstract}

\maketitle


Metal contacts to semiconductors must exhibit low resistance at short contact lengths for nanoscale electronics \cite{su_perspective_2022}. Since contact resistance increases exponentially with the Schottky barrier height $\Phi_B$, understanding and predicting $\Phi_B$ is central to contact physics. In the Schottky-Mott limit, the $n$-type barrier height is $\Phi_{Bn}=W_M-\chi_S$, where $W_M$ is the metal workfunction and $\chi_S$ is the semiconductor electron affinity \cite{schottky_zur_1939, mott_theory_1939}. This yields a pinning factor $S_p = \partial\Phi_{Bn}/\partial W_M=1$ and implies full tunability of $\Phi_B$ by varying $W_M$ \cite{tung_physics_2014}.

In practice, this limit is not attained due to Fermi-level pinning (FLP), where $S_p<1$ and $\Phi_{B}$ becomes insensitive to $W_M$. While FLP in conventional semiconductors is attributed to defect-induced gap states (DIGS) arising from dangling bonds of the semiconductor surface \cite{bardeen_surface_1947, cowley_surface_1965, tersoff_schottky_1984}, experiments \cite{sotthewes_universal_2019, kim_fermi_2017, wang_origins_2023} and density-functional theory (DFT) simulations \cite{gong_unusual_2014, fediai_towards_2016, guo_schottky_2015, kang_computational_2014} show that strong FLP occurs even in defect-free low-dimensional semiconductors \cite{novoselov_2d_2016}. This behavior is attributed to metal-induced gap states (MIGS), for which a comprehensive, predictive formalism has yet to be established \cite{tung_physics_2014}.

\begin{figure}[t!]
    \centering
    \includegraphics[width=\linewidth]{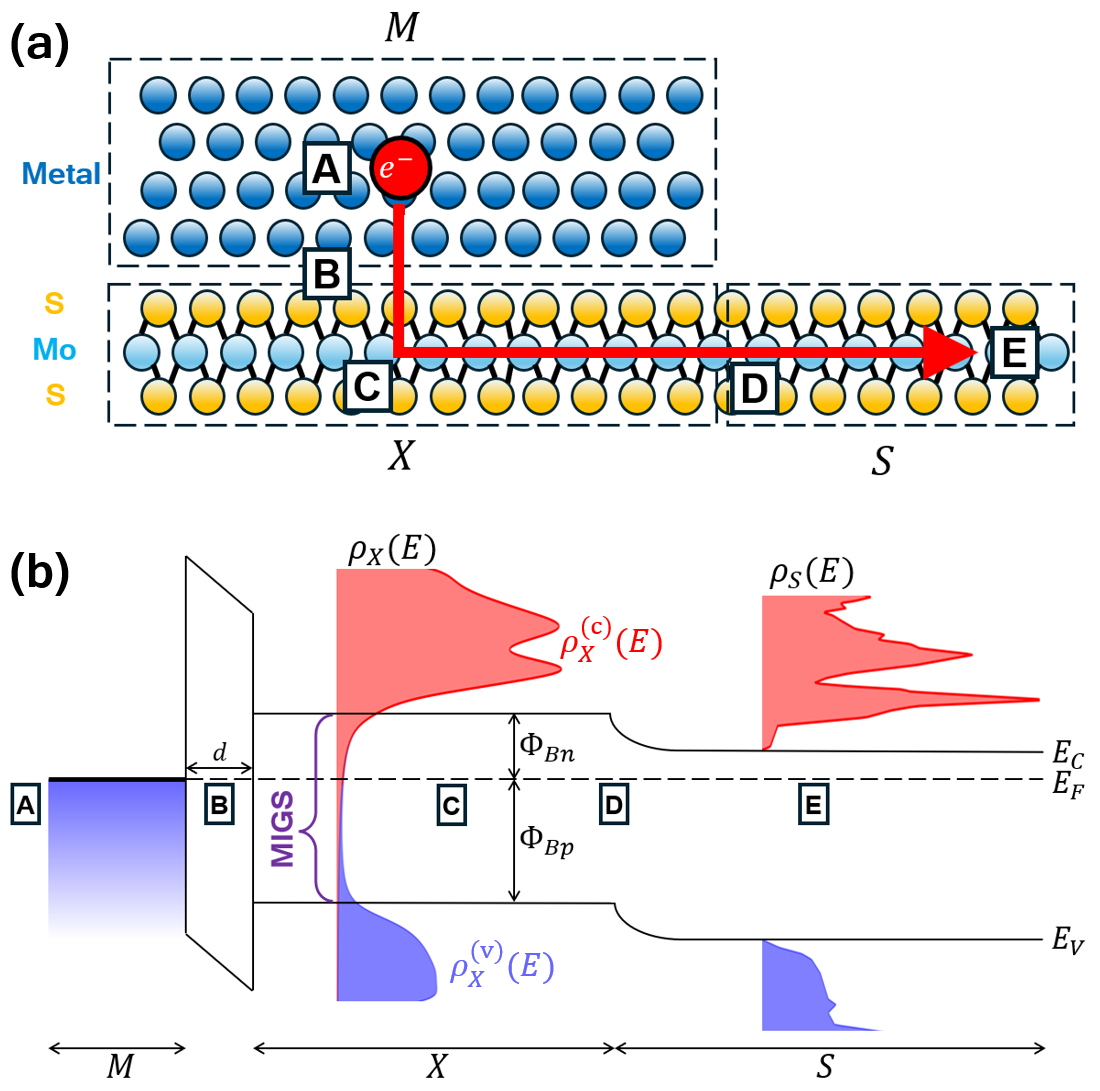}
    \caption{Schematic of a metal-on-TMD contact ($M$: metal, $S$: semiconductor, $X$: semiconductor-under-metal). Based on a figure from \cite{kang_computational_2014}. (a) Carriers follow the injection path A-B-C-D-E. (b) Energy band diagram along the injection path. The metal-semiconductor gap has thickness $d$. MIGS with density $\rho_X(E)$ arise in the bandgap, comprised of conduction-($\rho_X^\text{(c)}$, red) and valence-($\rho_X^\text{(v)}$, blue) band contributions. $n$- and $p$-type Schottky barrier heights are $\Phi_{Bn}$, $\Phi_{Bp}$.}
    \label{fig:Theory}
\end{figure}

In this Letter, we present a unified theory of MIGS and FLP in any dimension, reducing the determination of key observables---$\Phi_B$, transfer length, and contact resistance---to analytical equations in terms of fundamental material parameters. While applicable to arbitrary band structures, in the effective-mass approximation our method reveals a universal scaling law $\Delta^{n/2-2}$ for MIGS density, where $\Delta$ is the energy depth into the gap and $n$ is the dimensionality. Our model yields transparent interpretations of experimental results for carbon nanotube (CNT) and transition-metal dichalcogenide (TMD) contacts and provides a tractable alternative to numerical `black-box' simulations. This work thus offers a clearer understanding of contact physics, the associated trade-offs, and pathways to scalable, low-resistance interfaces. 

\emph{Analytical model of contact physics}---Let $M, S$, and $X$ denote the metal, semiconductor away from contact, and semiconductor-under-metal regions, respectively (\hyperref[fig:Theory]{Fig.~1(a)}), with corresponding densities of states (DoS) $\rho_{M,S,X}(E)$. Assuming a weak metal-semiconductor coupling characteristic of van der Waals interfaces \cite{farmanbar_first-principles_2016, ma_van_2024}, an electron of energy $E$ tunnels from $X$ to $M$ with rate $\Gamma(E)/\hbar$ given by Fermi's Golden Rule \cite{bardeen_tunnelling_1961, nemec_contact_2006, gottlieb_bardeens_2006}:
\begin{equation} \label{eq:Gamma}
    \Gamma(E) = 2\pi |t|^2 \rho_M(E).
\end{equation}
$t$ is the Hamiltonian matrix element between the wavefunctions of $M$ and $X$, which is approximately energy-independent \cite{bardeen_tunnelling_1961}. Since the electron has a finite lifetime $\hbar/\Gamma(E)$, its energy level broadens into a Lorentzian lineshape \cite{vazquez_dipole_2004, vazquez_energy_2005,mahan_many-particle_2000}. Every electron in $X$ experiences a similar interaction, so
$\rho_X(E)$ is obtained by convolving $\rho_S(E)$ with a Lorentzian kernel of width $\Gamma(E)$:
\begin{equation} \label{eq:convolution}
    \rho_X(E) = \int_{-\infty}^\infty \frac{\Gamma(E)/\pi}{(E'-E)^2 + |\Gamma(E)|^2} \rho_S (E') \d E'.
\end{equation}
MIGS density is defined as $\rho_X(E)$ in the bandgap. Thus, MIGS are Lorentzian tail states of the semiconductor bands resulting from energy broadening (\hyperref[fig:Theory]{Fig.~1(b)}), and can be split into conduction- and valence-band contributions since $\rho_S(E)$ is nonzero only in the bands:
\begin{equation} \label{eq:approxMIGS}
    \rho_X(E) = \rho_X^{\text{(c)}}(E)+\rho_X^{\text{(v)}}(E). 
\end{equation}
Deep in the bandgap ($\Delta \gg \Gamma$), $\rho_X^{\text{(c,v)}}(E)$ scales with the energy depth $\Delta$ from its respective band edge as
\begin{equation} \label{eq:approxMIGS_scale}
    \rho_X^{\text{(c,v)}}(E) \propto \Gamma(E)m_\text{eff}^{n/2}\Delta^{n/2-2}
\end{equation}
for an $n$-dimensional semiconductor with effective mass $m_\text{eff}$. We derive this scaling law and give exact formulas for $\rho_X^{\text{(c,v)}}(E)$ in \hyperref[eq:DoS_n]{Appendix A}. For a given coupling $\Gamma$, \cref{eq:approxMIGS_scale} suggests that lower-dimensional materials may exhibit inherently weaker FLP because their DoS provides fewer states to redistribute into the gap. Thus, MIGS is lowest in 1D (CNTs), followed by 2D (TMDs), and then 3D (\cref{fig:DoS_Dimensions}).

\begin{figure*}[t!]
    \centering
    \includegraphics[width=\linewidth]{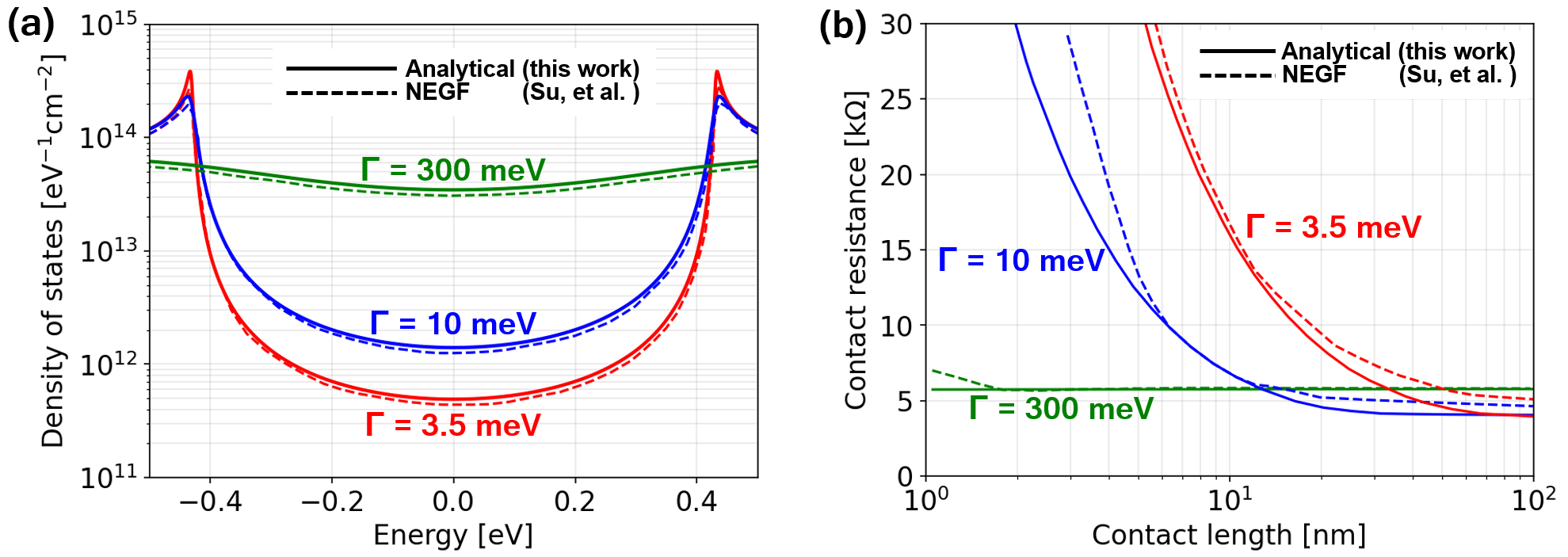}
    \caption{Validation of our analytical model against NEGF simulations by Su\emph{ et al.} \cite{su_effect_2023}. Analytical results (solid lines) show excellent agreement with NEGF results (dashed lines) across both MIGS density and length-dependence of contact resistance, $R_C(L_C)$. (a) MIGS density for Pd contact to $(13,0)$ CNT, at various coupling strengths $\Gamma$. The metal is assumed to be wideband \cite{covito_transient_2018}, so $\Gamma(E) = \text{const}$. van Hove singularities are visible at $E = \pm 0.43$ eV. (b) Contact resistance as a function of contact length and coupling $\Gamma$ for $p$-type contact, assuming $\Phi_{Bp}=0$ for Pd-on-CNT as in \cite{su_effect_2023}.}
    \label{fig:NEGF_CNT}
\end{figure*}

With $\rho_X(E)$ established, we incorporate it into the Cowley-Sze framework \cite{cowley_surface_1965} to predict the Schottky barrier height at the $M$--$X$ interface. Following V\'asquez \emph{et al.} \cite{vazquez_dipole_2004}, the charge neutrality level (CNL) of $X$ follows by requiring its total electron density to equal that of $S$:
\begin{equation} \label{eq:CNL}
    \int_{-\infty}^{\text{CNL}} \rho_X(E) \d E=\int_{-\infty}^{E_V} \rho_S (E) \d E.
\end{equation}
We discuss the relation between this definition of CNL and Tersoff's classic definition \cite{tersoff_schottky_1984} in \hyperref[eq:KK]{Appendix B}. In equilibrium, the net charge density on $X$ resulting from the displacement of the Fermi level $E_F$ from CNL is balanced by opposite charge on the metal surface. Cowley and Sze's model \cite{cowley_surface_1965} yields
\begin{equation} \label{eq:Fermilevel}
    \chi_S - W_M = E_F - E_C + \frac{q^2d}{\epsilon}\int_{\text{CNL}}^{E_F} \rho_X (E) \d E.
\end{equation}
The Schottky barrier heights are $\Phi_{Bn} = E_C-E_F$ and $\Phi_{Bp} = E_F-E_V$.

Macroscopic transport properties are obtained via thermal averages at the Fermi surface. The specific contact resistivity $\rho_C$, quantum resistance $R_Q$, and sheet resistance $R_\text{sh}$ of the $M$--$X$ interface are given by \cite{lundstrom_near-equilibrium_2013}
\begin{align}
        \rho_C^{-1} &= \frac{\pi q^2}{h} \int \d E \left(-\frac{\partial f_0}{\partial E}\right) \Gamma(E)\rho_X(E) \label{eq:contactresistivity}\\
        R_Q^{-1} &= \frac{2q^2}{h} \int \d E \left(-\frac{\partial f_0}{\partial E}\right) M(E)\label{eq:quantumresistance} \\
        R_\text{sh}^{-1} &= \frac{2q^2}{h} \int \d E \left(-\frac{\partial f_0}{\partial E}\right) M_\text{2D}(E)\lambda(E) \label{eq:sheetresistance}
\end{align}
where $f_0(E) = \frac{1}{\exp \big(\frac{E-E_F}{kT}\big)+1}$ is the Fermi-Dirac function, $M(E)$ is the number of transport modes in the semiconductor at energy $E$, $M_\text{2D}(E)$ is the mode density per unit width, and $\lambda(E)$ is the mean free path for backscattering in $X$. The exponential dependence of contact resistance on barrier height, $R_Q\propto \exp(\Phi_B/kT)$, emerges from \cref{eq:quantumresistance} and reflects thermal emission over the Schottky barrier.

Finally, these transport properties can be used to compute the transfer length $L_t$ and contact resistance $R_C$ at contact length $L_C$. We use the model of Solomon \cite{solomon_contact_2011}, repeated here for convenience, which generalizes the transmission-line model \cite{berger_models_1972} to ballistic (e.g., CNT) and diffusive (e.g., TMD) transport under the contact:
\begin{align}
    L_t &= \frac{1}{\sqrt{2\alpha\beta +\beta^2}} \label{eq:Lt} \\
    R_C &= \frac{R_Q}{2\beta L_t} \coth\bigg(\frac{L_C}{L_t}\bigg) \label{eq:Rc}
\end{align}
where $\alpha = \frac{R_\text{sh}}{R_Q}$ and $\beta = \frac{R_Q}{2\rho_C}$. \cref{eq:Lt} yields scaling laws for $L_t$ in terms of $\Gamma$ and other variables. Using $\langle \rangle$ to denote a thermal average, we find 
\begin{equation}
    L_t \sim \frac{v_\text{th}}{\langle \Gamma \rangle/\hbar} \label{eq:Lt_ballistic}
\end{equation}
in the ballistic limit $\langle \lambda \rangle \gg \hbar v_\text{th}/\langle \Gamma \rangle$, and 
\begin{equation}
    L_t \sim \sqrt{\frac{v_\text{th}\langle \lambda \rangle}{\langle \Gamma \rangle/\hbar}} \label{eq:Lt_diffusive}
\end{equation}
in the diffusive limit $\langle \lambda \rangle \ll \hbar v_\text{th}/\langle \Gamma \rangle$, where $v_\text{th}$ is the thermal velocity. \cref{eq:Lt_diffusive} is equivalent to the standard result $L_t=\sqrt{\rho_C/R_\text{sh}}$ \cite{berger_models_1972}. These scaling laws yield estimates of the interaction strength $\langle \Gamma \rangle$ from experimental values of $L_t$, regardless of the Schottky barrier height.

This model constitutes a complete and unified description of MIGS (\cref{eq:convolution,eq:approxMIGS}), band alignment (\cref{eq:CNL,eq:Fermilevel}), charge injection (\crefrange{eq:contactresistivity}{eq:sheetresistance}), transfer length (\cref{eq:Lt,eq:Lt_ballistic,eq:Lt_diffusive}), and overall contact resistance (\cref{eq:Rc}) in low-dimensional contacts.

\emph{Benchmarking against NEGF}---We validate our analytical model via direct comparison (\cref{fig:NEGF_CNT}) with NEGF simulations of a Pd-on-CNT contact by Su\emph{ et al.} \cite{su_effect_2023}, who used a coupled mode-space NEGF solver with $k\cdot p$ Hamiltonian. To ensure direct comparison, we adopt their assumptions: (1) $(13,0)$-chirality CNT (2) wideband metal, or $\Gamma(E)=\text{const}.$ \cite{covito_transient_2018} (3) $p$-type contact with zero Schottky barrier, inspired by low $\Phi_{Bp}$ observed in Pd-on-CNT experiments \cite{pitner_low-temperature_2019}.

The benchmarking results confirm the model's validity in two key ways. First, the nearly exact matching of MIGS density in \hyperref[fig:NEGF_CNT]{Fig.~2(a)} validates our treatment of MIGS as Lorentzian tail states and supports the $\Delta^{-3/2}$ scaling law (\cref{eq:approxMIGS_scale}) for 1D materials. Second, the excellent agreement for $R_C(L_C)$ in \hyperref[fig:NEGF_CNT]{Fig.~2(b)} verifies our model for $R_C$ and $L_t$ as a function of $L_C$ and $\Gamma$, confirming that stronger coupling enables shorter contacts. Although this comparison assumes a wideband metal, our analytical framework remains valid for more realistic $\Gamma(E)$ that violates the wideband assumption (\hyperref[eq:DoS_n]{Appendix A}).

\emph{Generalizability and implications}---Since the bandgap and electron affinity of CNTs are determined by their chirality, $\Phi_B$ exhibits a corresponding chirality dependence. Using material parameters obtained from \emph{ab initio} literature, our model predicts a ``pinning factor" of $\partial\Phi_{Bp}/\partial d_{\text{CNT}} \approx -0.4 \text{ eV}/\text{nm}$ for Pd-on-CNT contacts, where $d_\text{CNT}$ is the CNT diameter. This result is in excellent agreement with the DFT study of Tang\emph{ et al.} \cite{tang_band_2025}, showing that our analytical framework captures the chirality dependence of metal-on-CNT contacts without adjustable fitting parameters. Details are provided in the Supplemental Material \cite{MyPaper2026}.

Beyond 1D nanotubes, our framework provides a physical basis for the FLP observed in 2D TMDs \cite{gong_unusual_2014, kim_fermi_2017, wang_mechanism_2024} and explains that the shape of the MIGS density is key to the de-pinning of semimetal-on-TMD contacts \cite{shen_ultralow_2021, hoang_enabling_2024} because it shifts the CNL. Because TMD surfaces are often inherently defective, DIGS also contribute to FLP. We incorporate DIGS in our model by utilizing a semiconductor DoS $\rho_S(E)$ featuring defect-related peaks in the bandgap. Studies of metal- and semimetal-on-TMD contacts are detailed in the Supplemental Material \cite{MyPaper2026}.

Although we focused on 1D and 2D contacts, this model could be extended to conventional 3D contacts by accounting for the spatial decay of coupling into the bulk. In 3D, the Hamiltonian element $t(z)\propto e^{-\kappa z}$ \cite{heiman_effects_1965, lowell_tunnelling_1979} introduces a depth-dependent broadening $\Gamma(E,z)$, while the presence of dangling bonds (i.e., DIGS) implies substantial surface states $\rho_S(E,z=0)$ in the bandgap.

In summary, this Letter establishes a unified analytical framework for contacts in any dimension. Identifying MIGS as Lorentzian tail states of the semiconductor, we provide a transparent physical basis for FLP in low-dimensional interfaces, accurately reproducing rigorous simulation results without adjustable parameters. 
Our model uncovers the precise dependence of contact resistance on material parameters such as interface coupling, defectivity, dimensionality, and effective mass, offering a rigorous foundation for understanding and overcoming the contact bottleneck in low-dimensional electronics.

\emph{Acknowledgements}---We thank Gregory Pitner and Jack Evans for useful discussions. This work is supported in part by the Department of Defense, the Microelectronics Commons, the California-Pacific-Northwest AI Hardware Hub, and member companies of the Stanford SystemX Alliance.

\bibliography{zotero,manual}

\clearpage
\onecolumngrid
\begin{center}
    {\bfseries\large End Matter}
    \vspace*{\baselineskip}
\end{center}
\twocolumngrid

\appendix
\renewcommand{\theequation}{A\arabic{equation}}
\setcounter{equation}{0}
\renewcommand{\thefigure}{A\arabic{figure}}
\setcounter{figure}{0}
\renewcommand{\theHfigure}{A\arabic{figure}}

\begin{figure*}[t!] 
    \centering
    \includegraphics[width=\linewidth]{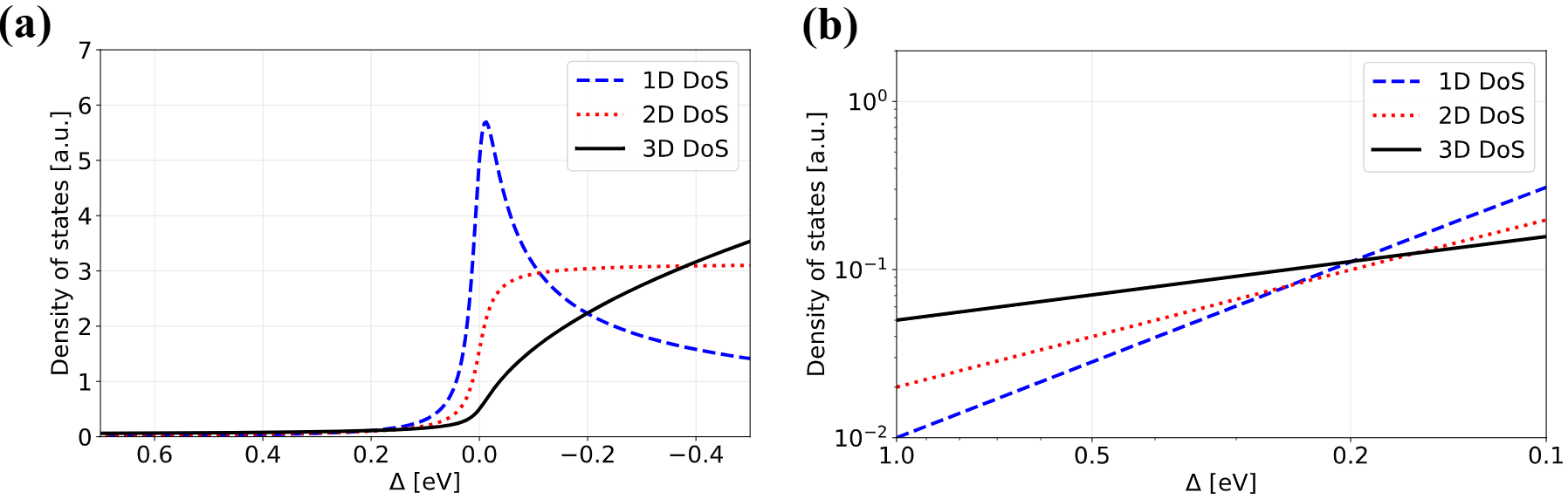}
    \caption{Dimensional dependence of conduction-band MIGS. We show the exact MIGS density $\rho_X^{(c)}$ in $n=1,2,3$ dimensions, calculated within the effective-mass approximation using \crefrange{eq:MIGS_1D}{eq:MIGS_3D}. The metal is assumed to be wideband with coupling strength $\Gamma = 20\text{ meV}$. $\Delta = E_C-E$ is the energy depth from the conduction band edge into the bandgap. (a) Linear-scale plot of the DoS. Interaction with the metal broadens the pristine band edges, giving rise to MIGS in the bandgap ($\Delta > 0$). (b) Log-log plot of MIGS density $\rho_X^\text{(c)}$ deep in the bandgap, demonstrating the universal $\Delta^{n/2-2}$ scaling law of \cref{eq:approxMIGS_scale}. \label{fig:DoS_Dimensions}}
\end{figure*}

\emph{Appendix A: MIGS density}---We derive \cref{eq:approxMIGS_scale} in the effective-mass approximation and provide general formulas valid anywhere in the bandgap. 
The DoS of the conduction band of an $n$-dimensional semiconductor is 
\begin{equation} \label{eq:DoS_n}
    \rho_S^\text{(c)}(E) = A(E-E_C)^{n/2-1}\Theta(E-E_C)
\end{equation}
where $A$ depends on effective mass, $A \propto m_\text{eff}^{n/2}$ \cite{lundstrom_near-equilibrium_2013}. To evaluate \cref{eq:convolution} for the conduction band, we use the Green's function formalism \cite{mahan_many-particle_2000, datta_electronic_1995}
\begin{equation} \label{eq:ImGz}
    \rho_X^\text{(c)}(E) = -\frac{1}{\pi} \text{Im}(G(z))
\end{equation}
with $z(E) = E-E_C+i\Gamma(E)$. The corresponding complex Green's functions in $n=1,2,3$ dimensions are \cite{economou_greens_2006}
\begin{align}
    G_\text{1D}(z) &= -A\pi/\sqrt{-z}\\
    G_\text{2D}(z) &= -A \ln(-z)\\
    G_\text{3D}(z) &= -A\pi \sqrt{-z} 
\end{align}
Defining energy depth into the gap as $\Delta(E) = E_C-E$, \cref{eq:ImGz} returns the following exact expressions for conduction-band MIGS, which modify \cref{eq:DoS_n} to account for interactions with a metal (\cref{fig:DoS_Dimensions}):
\begin{align}
    \rho_{X, \text{1D}}^\text{(c)}(E) &= A \sqrt{\frac{{\sqrt{\Delta^2+|\Gamma(E)|^2}}-\Delta}{2(\Delta^2+|\Gamma(E)|^2)}} \label{eq:MIGS_1D}\\
   \rho_{X, \text{2D}}^\text{(c)}(E) &= \frac{A}{\pi} \cot^{-1}\left(\frac{\Delta}{\Gamma(E)}\right) \label{eq:MIGS_2D}\\
    \rho_{X, \text{3D}}^\text{(c)}(E) &=  A \sqrt{\frac{\sqrt{\Delta^2+|\Gamma(E)|^2}-\Delta}{2}}\label{eq:MIGS_3D}
\end{align}
These expressions are valid even if the metal is not wideband ($\Gamma(E) \neq \text{const}$). Deep in the bandgap $\Delta \gg\Gamma(E)$, \cref{eq:MIGS_1D,eq:MIGS_2D,eq:MIGS_3D} simplify to
\begin{align}
    \rho_{X, \text{1D}}^\text{(c)}(E)&\approx \frac{A\Gamma(E)}{2}\Delta^{-3/2}\\
   \rho_{X, \text{2D}}^\text{(c)}(E) &\approx \frac{A\Gamma(E)}{\pi}\Delta^{-1}\\
   \rho_{X, \text{3D}}^\text{(c)}(E) &\approx \frac{A\Gamma(E)}{2}\Delta^{-1/2}
\end{align}
which results in \cref{eq:approxMIGS_scale}. However, for even deeper energies $\Delta \gg W$, where $W$ is the width of the conduction band, the entire band is indistinguishable from a single energy level. The DoS then decays as a Lorentzian tail $\rho_{X}^\text{(c)}\propto \Delta^{-2}$ regardless of dimension. One may use these approximations for $\rho_X^\text{(c,v)}(E)$ to estimate the pinning factor $S_p$ via the classic equation \cite{tung_physics_2014}
\begin{equation} \label{eq:pinfactor}
    S_p = \left(1+ \frac{q^2d}{\epsilon}\rho_X(E_F)\right)^{-1} .
\end{equation}

\renewcommand{\theequation}{B\arabic{equation}}
\setcounter{equation}{0}
\emph{Appendix B: Discussion of charge neutrality level}---We discuss how \cref{eq:CNL}, which defines CNL through explicit charge conservation, relates to Tersoff's well-known theory of CNL \cite{tersoff_schottky_1984}. 
The latter defines CNL as the solution in $E$ to $\text{Re }G(\mathbf R,E)=0$, where $G(\mathbf R,E)$ is the real-space Green's function of the pristine semiconductor and the distance $\mathbf R$ is large.
We present an intuitive understanding of Tersoff's formulation starting with the Kramers-Kr\"onig relation, using $\mathcal P$ to denote the principal value:
\begin{equation} \label{eq:KK}
    \text{Re }G(\mathbf R,E) = \mathcal P \int_{-\infty}^\infty \frac{\text{Im }G(\mathbf R,E)}{E-E'} \d E'.
\end{equation}
\cref{eq:ImGz} yields $\text{Im }G(\mathbf R,E)=-\pi \rho_S(\mathbf R,E)$, where $\rho_S(\mathbf R,E)$ is the number of states at energy $E$ whose size is at least $|\mathbf R|$. Since $\mathbf R$ is large, $\rho_S(\mathbf R,E)$ selects the states of the semiconductor that dominate the gap states because they protrude the furthest at the surface, making up the dangling bonds (DIGS) and interacting strongly with the metal (MIGS). 
$\text{Re }G(\mathbf R, \text{CNL}) = 0$ becomes
\begin{align} \label{eq:simpleTersoff}
    0=\int_{-\infty}^{\infty}  \frac{\rho_S(\mathbf R,E')}{\text{CNL}-E'} \d E'.
\end{align}
Since $\frac{1}{\text{CNL}-E'}$ is the integral of a Lorentzian tail with very small width $\frac{1}{(E'-E)^2}$, \cref{eq:simpleTersoff} is equivalent to
\begin{equation} \label{eq:Tersoff_CNL_X}
    \int_{-\infty}^{\text{CNL}}\rho_X(\mathbf R, E) \d E =\int_\text{CNL}^\infty \rho_X(\mathbf R, E) \d E
\end{equation}
where $\rho_X(\mathbf R, E)$ is the result of very small, energy-independent broadening applied to $\rho_S(\mathbf R, E)$:
\begin{equation}
        \rho_X(\mathbf R, E) = \lim_{\Gamma \rightarrow 0} \int \frac{\Gamma/\pi}{(E'-E)^2 + \Gamma^2} \rho_S (\mathbf R, E') \d E'.
\end{equation}
Because \cref{eq:CNL} and \cref{eq:Tersoff_CNL_X} are equivalent, we find that Tersoff's formulation correctly enforces charge conservation under the following conditions: (1) only the longest-ranged states dominate MIGS (2) these states experience weak, energy-independent broadening. Condition (2) is a reasonable assumption for the ``intrinsic" CNL of the semiconductor surface, since the metal is unknown.

In summary, we have shown that Tersoff’s formulation of CNL is a limiting case of the broader principle of charge conservation. By formulating CNL through explicit charge conservation rather than the zero of $G(\mathbf R,E)$, our framework remains valid even in the presence of energy-dependent coupling, or structural disorder where the Green's function $G(\mathbf R, E)$ of the periodic crystal no longer accurately describes the local interface phenomena that cause FLP.

\end{document}



\title{\textbf{Supplemental Material for \\ Contacts to Low-Dimensional Semiconductors: \\Physical Theory and Analytical Model}}


\author{Jimmy Qin}
\author{H.-S. Philip Wong}
\affiliation{%
Department of Electrical Engineering, Stanford University, Stanford, CA 94305, USA
}




\maketitle

This document provides detailed applications of our analytical theory \cite{MainPaper2026} to various contacts to low-dimensional semiconductors. We provide analyses of the following topics:

\begin{itemize}
    \item Dependence of the Schottky barrier height on the chirality of the carbon nanotube (CNT) in a metal-on-CNT contact. We use Pd-on-CNT as our case study.
    \item Dependence of the Schottky barrier height and pinning strength in metal-on-TMD contacts on TMD defectivity and metal-semiconductor coupling. We use Au-on-MoS$_2$ as our case study.
    \item Explanation of why semimetal-on-TMD contacts have weaker pinning than metal-on-TMD contacts. We show that the reshaped MIGS density of semimetal contacts leads to an improved pinning factor $S'=(2S+1)/3$ compared to the pinning factor $S$ of metal contacts with similar workfunction and coupling, which explains experimentally observed improvements in Schottky barrier height \cite{shen_ultralow_2021}.
\end{itemize}

\section{Metal-on-CNT contact: $\Phi_{B}$ vs. chirality} \label{sec:s1}
Chirality describes how a graphene sheet is rolled up into a nanotube and determines CNT properties important for contact physics, such as band gap $E_g$, diameter $d_\text{CNT}$, and density of states (DoS) \cite{wong_carbon_2011}. Understanding the effect of chirality on CNT contacts is important in choosing the optimal chirality for a CNT field-effect transistor due to the trade-off between low contact resistance and low device leakage \cite{chiu_overcoming_2025}. In this section, we study the effect of CNT chirality on the $p$-type Schottky barrier $\Phi_{Bp}$ of a Pd-on-CNT contact and compare our analytical results to density-functional theory (DFT) results obtained by Tang\emph{ et al.} \cite{tang_band_2025}.

To compute the density of states and bandgap of the CNT, we use the standard tight-binding formalism \cite{wong_carbon_2011} for CNTs with a carbon-carbon distance of $a = 1.42 \text{ \AA}$ and hopping energy $\gamma_0 = 2.7 \text{ eV}$, keeping only the leading-order contributions (for example, we use $E_g \propto d_\text{CNT}^{-1}$ and ignore the correction of order $d_\text{CNT}^{-2}$). Other quantities required for our model are obtained from theoretical or \emph{ab initio} simulation literature, described in \cref{tab:table1}. 

\begin{figure}[t]
    \centering  
    \includegraphics[width=\linewidth]{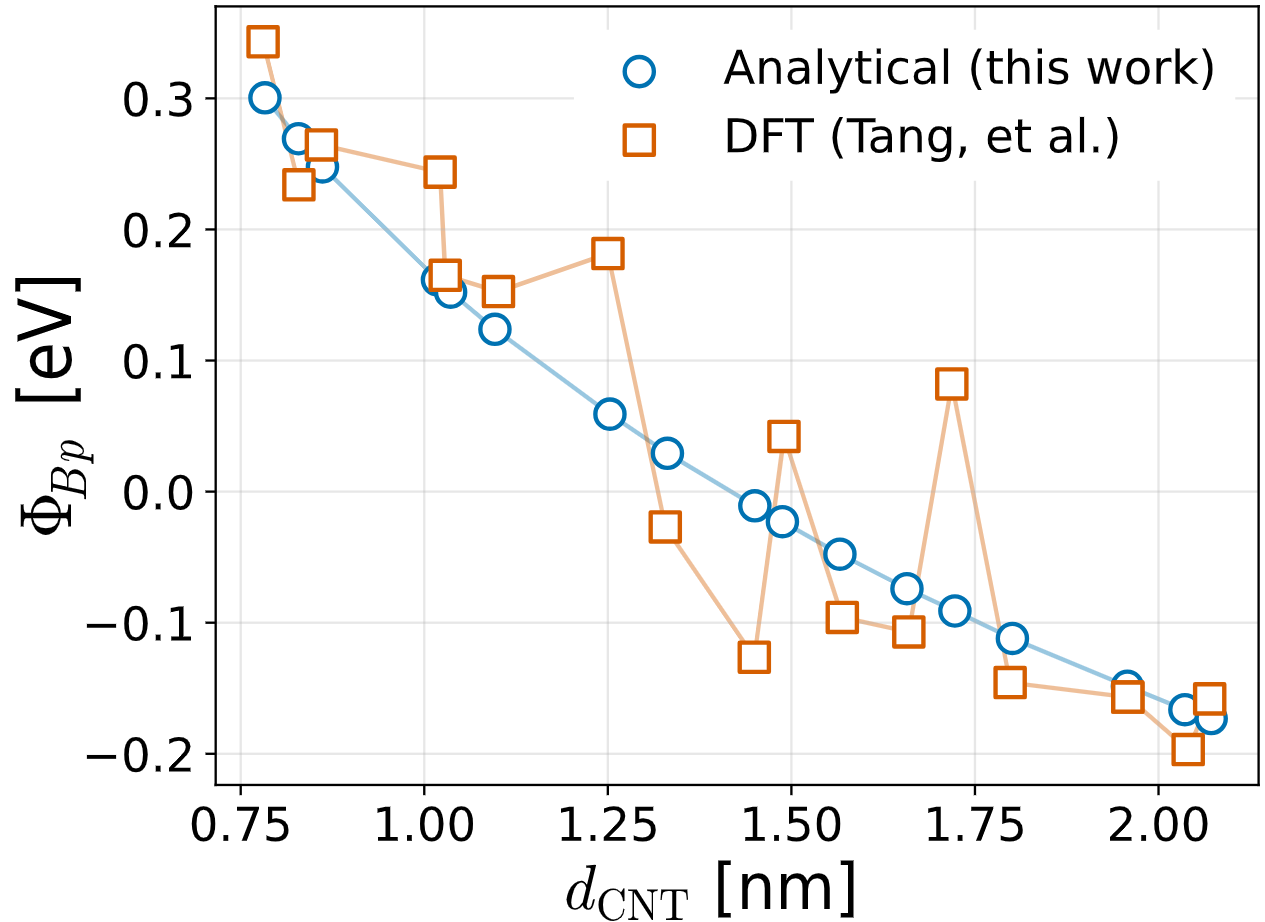}
    \caption{Dependence of $p$-type Schottky barrier height $\Phi_{Bp}$ on CNT chirality for a Pd-on-CNT contact. CNT diameter $d_\text{CNT}$ is a function of chirality. Each point corresponds to a CNT of different chirality (blue circles: our analytical model; orange squares: DFT simulation of Tang\emph{ et al}. \cite{tang_band_2025}). Our analytical model quantitatively reproduces the trend $\partial \Phi_{Bp}/\partial d_\text{CNT}\approx -0.4 \text{ eV}/\text{nm}$ observed in the DFT simulations. \label{fig:chirality}}
\end{figure}

Our results are shown in \cref{fig:chirality}, displaying the trend $\partial \Phi_{Bp}/\partial d_\text{CNT}\approx -0.4 \text{ eV}/\text{nm}$, which matches precisely with the DFT simulation results of Tang\emph{ et al.} \cite{tang_band_2025}. The DFT results appear `noisier' in \cref{fig:chirality} because the authors found significant dependence of $\chi_S$ on chirality even for CNTs of similar diameter. In contrast, we used an empirical formula for $\chi_S$ which depends only on $E_g$ (\cref{tab:table1}), which captures the primary trends of $\chi_S$ but smooths out chirality-related fluctuations. Even so, our model's quantitative reproduction of the main trend of $\Phi_B$ suggests it is well-suited for guiding experimental device design and performing analyses without the prohibitive cost of atomistic simulations. Another advantage of our approach is the ease of investigating hypothetical scenarios. For example, if the metal-CNT interaction is weaker in reality than DFT predicts, our model provides a simple way to compute the effect of this discrepancy on contact performance.

\begin{table*}
\caption{\label{tab:table1}Parameters used in analytical modeling of Pd-on-CNT contacts}
\setlength{\tabcolsep}{8pt}

\begin{tabular}{c c cc}
\hline\hline
Quantity & Symbol & Value & Notes \\
\hline

Pd--CNT distance
& $d$
& 3.5~\AA\ \cite{vanin_graphene_2010}
&\parbox[t]{0.3\textwidth}{
\setlength{\baselineskip}{12pt}Assuming Pd--CNT and Pd--graphene distances are similar.} 
\\

Pd--CNT interaction strength
& $\Gamma$
& $0.38\,\mathrm{eV}$ \cite{nemec_contact_2006}
& 
\\

Work function of Pd
& $W_M$
& $5\,\mathrm{eV}$ \cite{gay_electronic_1981}
& 
\\

Dielectric constant of CNT
& $\epsilon$
& $1+\left(\frac{\hbar\omega_p}{5.4\,E_g}\right)^2$ \cite{krupke_separation_2003}
& \parbox[t]{0.3\textwidth}{
\setlength{\baselineskip}{12pt}Penn model \cite{penn_wave-number-dependent_1962} for dielectric constant. $\hbar\omega_p \approx 5\,\mathrm{eV}$ is the plasma frequency along the CNT axis.}
\\

Electron affinity of CNT
& $\chi_S$
& $W_M^{\mathrm{Gr}}-\frac{E_g}{2}$ \cite{akturk_quantum_2005}
& \parbox[t]{0.3\textwidth}{
\setlength{\baselineskip}{12pt}$W_M^{\mathrm{Gr}} = 4.5\,\mathrm{eV}$ is the work function of graphene.}
\\

\hline\hline
\end{tabular}

\end{table*}

\section{Metal-on-TMD contact: Defectivity and pinning}
\label{sec:s2}
While the main text \cite{MainPaper2026} and \cref{sec:s1} focused on 1D CNTs, our theory is equally applicable to 2D TMDs. In metal-on-TMD contacts, Fermi-level pinning (FLP) is caused by both metal-induced gap states (MIGS) and defect-induced gap states (DIGS), due to the defectivity of the TMD \cite{bampoulis_defect_2017,mcdonnell_defect-dominated_2014,liu_sulfur_2013,huang_role_2017}. For example, the dominant defect species in monolayer MoS$_2$ is the sulfur vacancy, denoted $\text{V}_\text{S}$, which pins the Fermi level near its trap energy, about $0.3 \text{ eV}$ below the conduction band minimum (CBM) \cite{liu_sulfur_2013, gali_electronic_2020}.

Despite the efforts devoted to improving TMD contacts, a clear assessment of the relative contributions of MIGS and DIGS to FLP is currently lacking \cite{liu_fermi_2022}. In this section, we demonstrate that our analytical model can fill this gap. We choose Au-on-MoS$_2$ for this study, since Au is a common metal for MoS$_2$ contacts \cite{english_improved_2016}. Using parameters from experimental and \emph{ab initio} simulation literature (\cref{tab:table2}), we utilize our analytical model to compute band alignment quantities---charge neutrality level CNL, $\Phi_{Bn}$, pinning factor $S_p$---for wide ranges of metal-semiconductor coupling $\Gamma(E)$ and sulfur vacancy concentration $[\text{V}_\text{S}]$. For simplicity, the metal is assumed to be wideband, so $\Gamma(E) = \Gamma$ is independent of $E$ \cite{covito_transient_2018}.

\begin{figure*}[t]
    \centering  
    \includegraphics[width=\linewidth]{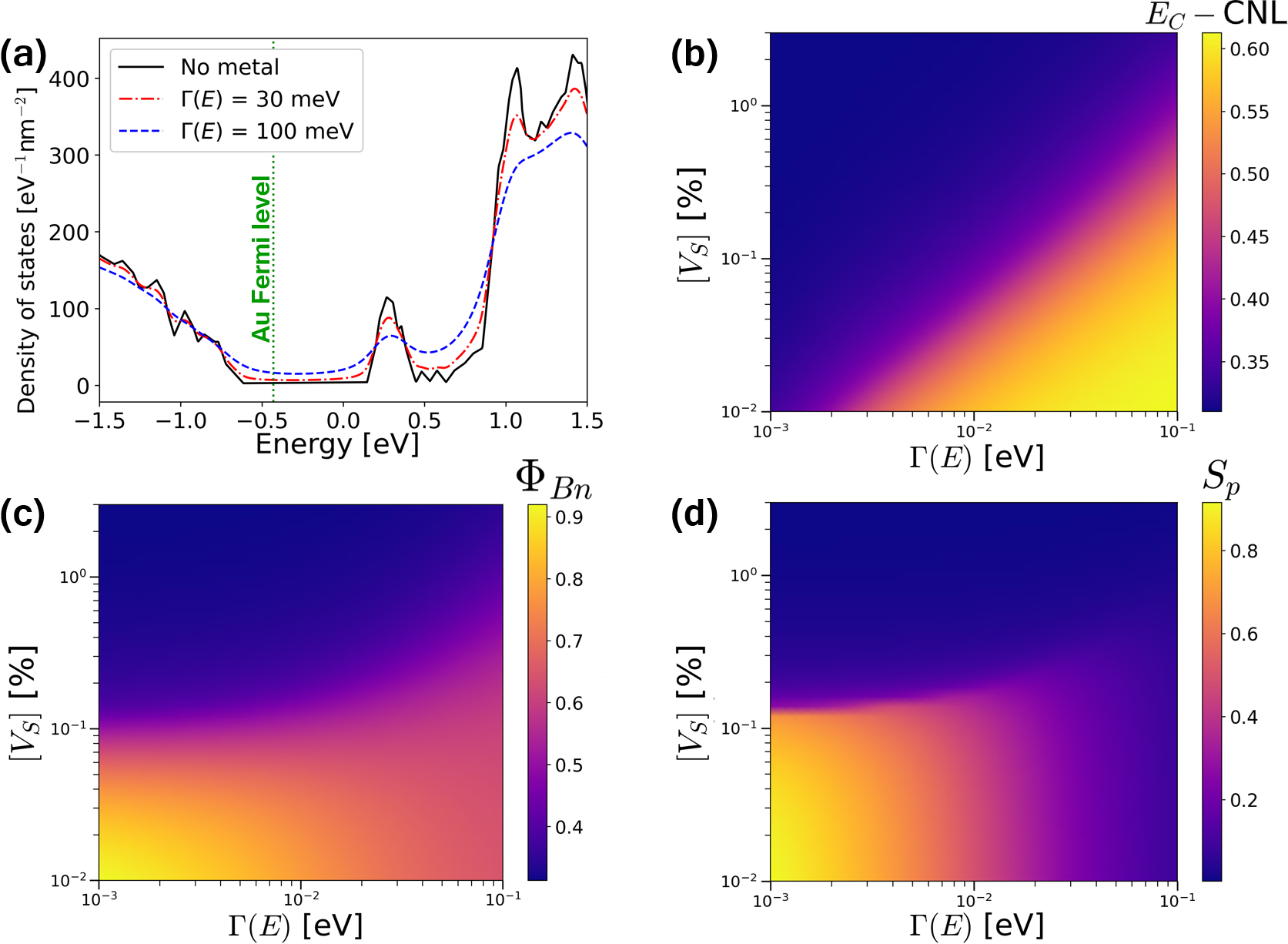}
    \caption{\label{fig:AuMoS2}Influence of metal coupling (MIGS) and defectivity (DIGS) on Au-on-$\text{MoS}_2$ contacts.
(a) DoS of monolayer $\text{MoS}_2$ with $[\text{V}_{\text{S}}]=3\%$ sulfur vacancy concentration and varying coupling strengths $\Gamma(E)$, from \cite{gali_electronic_2020}. The vacancy peak is centered at $E \approx 0.3 \text{ eV}$, about $0.3 \text{ eV}$ below the CBM, and the Fermi level of Au is shown.
(b) Calculated CNL relative to the CBM as a function of $\Gamma(E)$ and $[\text{V}_{\text{S}}]$. $\text{CNL}$ transitions from the intrinsic limit \cite{tersoff_schottky_1984} ($\sim 0.61 \text{ eV}$ below CBM) to the defect-dominated limit ($\sim 0.31 \text{ eV}$ below CBM).
(c) $n$-type Schottky barrier height $\Phi_{Bn}$, showing the transition from the Schottky-Mott limit ($\Phi_{Bn}\sim 1 \text{ eV}$) to DIGS-induced pinning ($\Phi_{Bn}\sim 0.3 \text{ eV}$).
(d) Pinning factor $S_p=\partial\Phi_{Bn}/\partial W_M$; lower values indicate stronger pinning driven by high MIGS (right) or high DIGS (top).}
\end{figure*}

Our results are shown in \cref{fig:AuMoS2}. The interplay between MIGS and DIGS reveals distinct regimes of pinning. In the low-coupling, low-defect limit (lower-left of \hyperref[fig:AuMoS2]{Fig.~S2(b,c,d)}), the interface approaches the Schottky-Mott limit, where $S_p = 1$ and $\Phi_{Bn} = W_M-\chi_S$. However, the introduction of even a modest sulfur vacancy concentration ($[V_{\text{S}}] > 0.1\%$) rapidly shifts the CNL toward the sulfur defect peak, pinning the Schottky barrier at $\Phi_{Bn} \sim 0.3\text{ eV}$ regardless of metal coupling strength (\hyperref[fig:AuMoS2]{Fig.~S2(c)}). In the high-coupling, low-defect limit, strong pinning occurs due to MIGS rather than DIGS (lower-right of \hyperref[fig:AuMoS2]{Fig.~S2(d)}).

This analysis enables a quantitative understanding of the relative contributions of MIGS and DIGS in TMD contacts. To understand the contribution of MIGS, we infer $\Gamma$ from experimental data using Eq. (13) in the main text, which describes how the transfer length $L_t$ scales with $\Gamma$:
\begin{equation}
    L_t \sim \sqrt{\frac{v_\text{th}\langle \lambda \rangle}{\langle \Gamma \rangle/\hbar}} \label{eq:Lt_diffusive}
\end{equation}
where $\langle \rangle$ denotes a thermal average. Using the values in \cref{tab:table2}, we find a very weak interaction strength $\langle \Gamma \rangle \approx 0.2 \text{ meV}$. According to \cref{fig:AuMoS2}, $\Gamma \approx 0.2 \text{ meV}$ suggests that MIGS have almost no impact in Au-on-MoS$_2$ contacts, at least for a transfer length of $L_t = 35\text{ nm}$ \cite{english_improved_2016}; we conclude that the energy band alignment in Au-on-MoS$_2$ contacts is dominated by DIGS. This logic can be reversed: short transfer length $L_t$ is an experimental signature of high MIGS.

To evaluate the impact of $\Gamma$ on contact resistance, we consider the scaling of $R_C$ with $\Gamma$. According to Eq. (11) of the main text \cite{MainPaper2026}, $R_C$ depends on $\Gamma$ as
\begin{equation}
    R_C(\Gamma) \sim \Gamma^{-1} e^{\Phi_B(\Gamma)/kT}
\end{equation}
where $\Gamma^{-1}$ and $\Phi_B(\Gamma)$ describe the dependence of specific contact resistivity $\rho_C$ and barrier height on $\Gamma$, respectively. Thus, $\Gamma$ should be optimized to minimize $\rho_C$ while remaining below the threshold where MIGS-induced FLP becomes significant---identified as $\Gamma \sim \text{3 meV}$ from \cref{fig:AuMoS2}. Given the estimate of $\Gamma \sim 0.2 \text{ meV}$ for Au-on-MoS$_2$ contacts, experimental efforts should \emph{decrease defectivity} (to reduce FLP) and \emph{increase metal coupling strength} (to reduce $\rho_C$). 
While these findings address the specific Au-MoS$_2$ interface, our model establishes a general framework to determine the optimal coupling for any metal-TMD combination. By integrating experimental $L_t$ and $\Phi_B$ measurements, it provides a quantitative diagnostic to assess whether contact performance is fundamentally limited by intrinsic metal-semiconductor coupling (MIGS) or extrinsic fabrication defects (DIGS).

\begin{table*}
\caption{\label{tab:table2}Parameters used in analytical modeling of Au-on-MoS$_2$ contacts}
\setlength{\tabcolsep}{8pt}
\begin{tabularx}{\textwidth}{cccc}
        \hline \hline
        Quantity & Symbol & Value & Notes  \\
        \hline
        \parbox[t]{0.3\textwidth}{
\setlength{\baselineskip}{12pt}
Density of states
of monolayer MoS$_2$
} & $\rho_S(E)$ & Obtained from \cite{gali_electronic_2020} & \parbox[t]{0.3\textwidth}{
\setlength{\baselineskip}{12pt}
Computed DoS in \protect\cite{gali_electronic_2020} covers a range of sulfur vacancy concentrations, $0\le [\text{V}_\text{S}] \le 3\%.$
}\\
        Workfunction of Au & $W_M$ & 5.2 eV \cite{ishida_work_2020}  &   \\
        \parbox[t]{0.3\textwidth}{
\setlength{\baselineskip}{12pt}Electron affinity of monolayer MoS$_2$} & $\chi_S$ & 4.2 eV \cite{xiao_enhanced_2018} &   \\
        Au-MoS$_2$ distance & $d$ & 3.5 \AA $\text{ }$ \cite{huang_large-period_2025} &   \\
       \parbox[t]{0.3\textwidth}{
\setlength{\baselineskip}{12pt}Dielectric constant of monolayer MoS$_2$} & $\epsilon$ & 3 \cite{belete_dielectric_2018} & \parbox[t]{0.3\textwidth}{
\setlength{\baselineskip}{12pt}In the perpendicular direction to the MoS$_2$ plane.} \\
        \parbox[t]{0.3\textwidth}{
\setlength{\baselineskip}{12pt}Transfer length of MoS$_2$ contacts} & $L_t$ & 35 nm \cite{english_improved_2016} &   \\
        Thermal velocity & $v_\text{th}$ & $2\times 10^5 \text{ m/s}$ & \parbox[t]{0.3\textwidth}{
\setlength{\baselineskip}{12pt}Estimated using electron effective mass of $0.35 m_e$ \protect\cite{cheiwchanchamnangij_quasiparticle_2012}.}  \\
        \parbox[t]{0.3\textwidth}{
\setlength{\baselineskip}{12pt}Mean free path for backscattering in MoS$_2$} & $\lambda$ & 2 nm \cite{english_approaching_2016} &   \\

         \hline \hline
\end{tabularx}
\end{table*}

\section{Semimetal-on-TMD contact: explanation of de-pinning}
\label{sec:s3}
Semimetal-on-TMD contacts have emerged as an effective method to reduce $\Phi_B$ and improve contact resistance $R_C$ \cite{liu_contact_2025}. Our framework shows that this improvement in $\Phi_B$ is not driven by a simple reduction in MIGS, but rather by an energy-dependent redistribution of MIGS that shifts the charge neutrality level (CNL) toward the relevant band edge.

\begin{figure*}[t]
    \centering  
    \includegraphics[width=\linewidth]{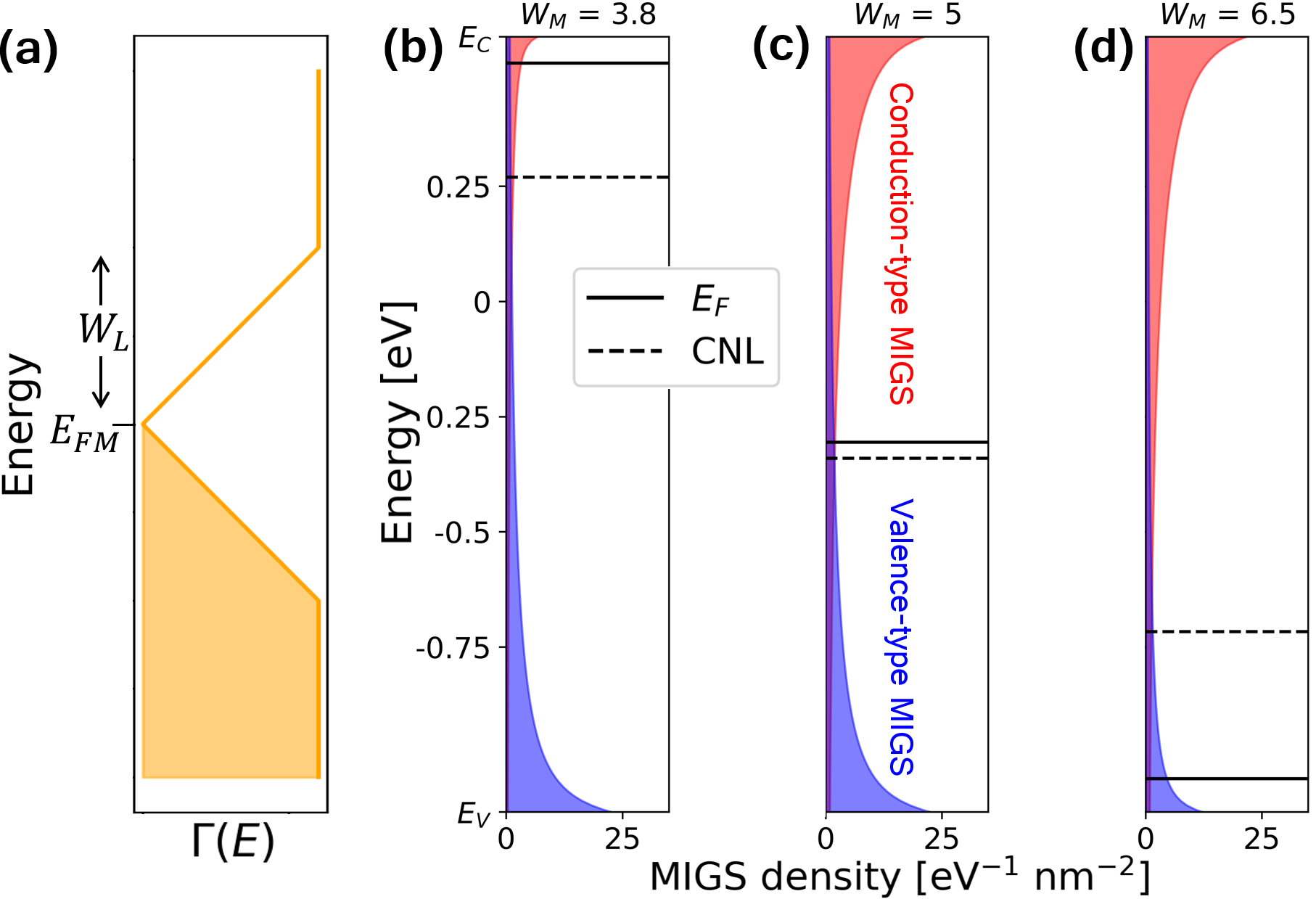}
    \caption{\label{fig:semimetal}Mechanism of Fermi level de-pinning in semimetal-on-TMD contacts. (a) Energy-dependent coupling strength $\Gamma(E)$ of a semimetal with linear dispersion region of width $W_L$ centered at $E_{FM}$ (\cref{eq:semimetal_Gamma}). Orange shading indicates occupied semimetal states. (b-d) Computed MIGS density in defect-free MoS$_2$ for semimetal workfunctions $W_M = 3.8, 5.0$, and $6.8 \text{ eV}$, respectively. Red and blue shading denote conduction- and valence-band contributions to MIGS. Dashed lines indicate CNL, while solid lines show equilibrium Fermi level $E_F$. Changing $E_{FM}$ (or equivalently $W_M =E_\text{vac}-E_{FM}$) pulls CNL towards $E_{FM}$, facilitating low-barrier $n$-type (b) and $p$-type (d) contacts. These plots were generated using the full integral formalism for band alignment (see the main text \cite{MainPaper2026}) rather than the analytical approximations developed in this section. The results of (b-d) confirm our analytical prediction of $\frac{\d \text{CNL}}{\d E_{FM}}\approx 1/3$ in the linear dispersion region.}
\end{figure*}

\begin{figure*}[t]
    \centering  
    \includegraphics[width=\linewidth]{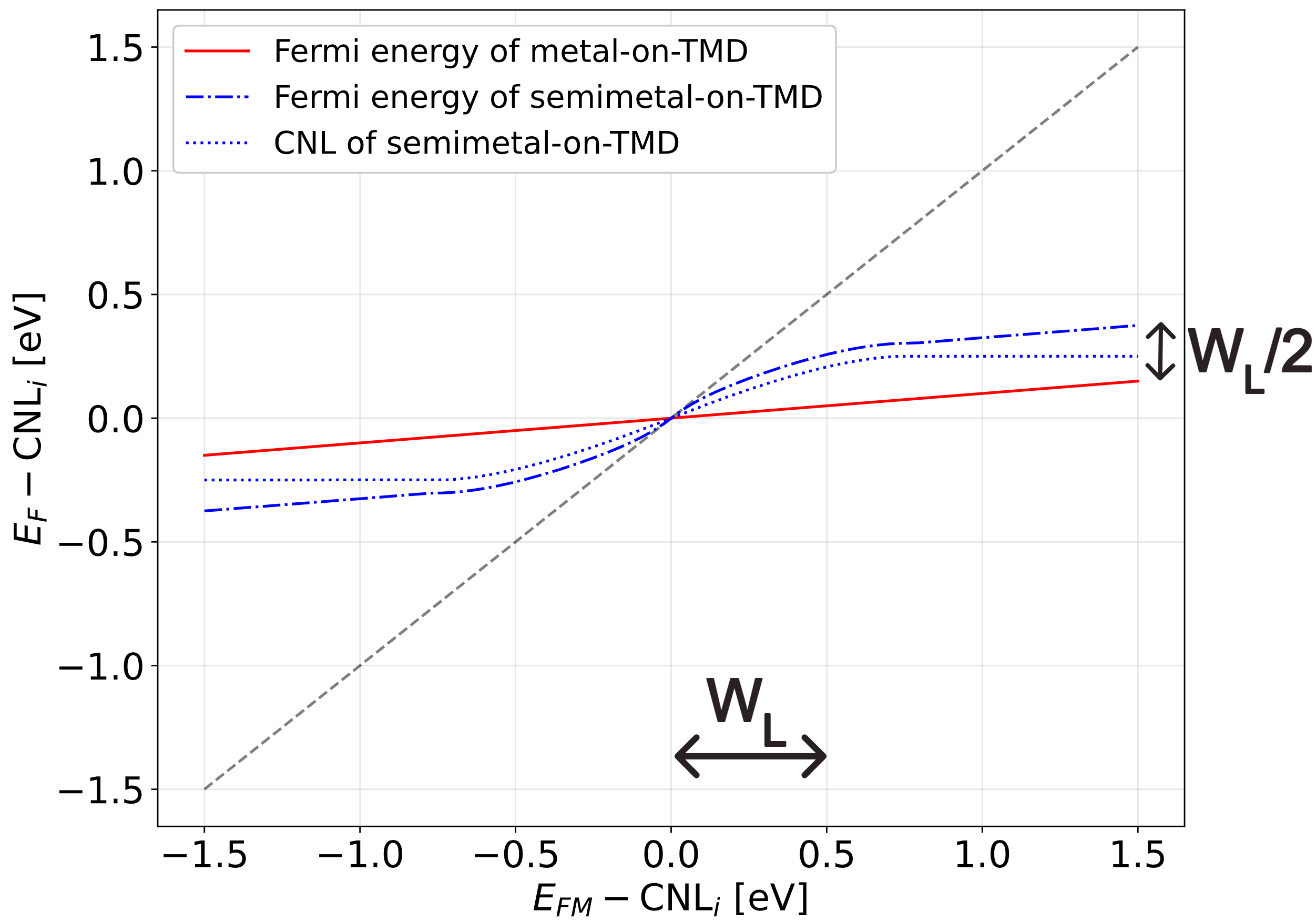}
    \caption{\label{fig:semimetal_2}Fermi level ($E_F$) and charge neutrality level ($\text{CNL}$) evolution as a function of the metal work function $E_{FM}$ for metal and semimetal contacts. $\text{CNL}_i$ is the charge neutrality level of the metal contact, which is independent of $E_{FM}$. The grey dashed line represents the unpinned Schottky-Mott limit ($S=1$), while the red line indicates a strongly pinned metal contact ($S=0.1$). The blue curves illustrate the de-pinning effect of a semimetal contact, where the $\text{CNL}$ shifts in response to the semimetal work function within the linear dispersion region $|E_{FM}-\text{CNL}_i|\lesssim W_L$, where $W_L = 0.5 \text{ eV}$ in this case. Beyond the linear dispersion region, $\text{CNL}$ saturates and the system reverts to standard metallic pinning behavior.} 
\end{figure*}

According to our framework \cite{MainPaper2026}, the MIGS density is proportional to the coupling $\Gamma(E)$, which itself scales with the metal DoS, $\rho_M(E)$. While a standard wideband metal provides a roughly constant coupling $\Gamma(E)\approx \Gamma_0$ \cite{covito_transient_2018}, semimetals exhibit a linear dispersion near their Dirac or Weyl points \cite{yan_topological_2017} and have vanishing DoS at the Fermi level $E_{FM}$ (\hyperref[fig:semimetal]{Fig.~S3(a)}). Consequently, when the semimetal $E_{FM}$ is aligned near the semiconductor CBM $E_C$, the conduction-band states experience significantly less broadening compared to the valence-band states. This asymmetry causes the valence-band contribution to dominate the MIGS, effectively ``pushing" the CNL toward $E_C$ (\hyperref[fig:semimetal]{Fig.~S3(b-d)}). The system Fermi level $E_F$ is pinned close to CNL, so this CNL shift effectively de-pins $E_F$ from the intrinsic charge neutrality level $\text{CNL}_i$ of the semiconductor (\cref{fig:semimetal_2}).

We formalize this by deriving approximate expressions for CNL, $\Phi_B$, and pinning factor in semimetal-on-TMD contacts. We utilize a coupling profile that accounts for the semimetal's linear dispersion in a energy range of width $W_L$ (\hyperref[fig:semimetal]{Fig.~S3(a)}), modifying the wideband approximation $\Gamma(E) \approx \Gamma_0$ for a regular metal \cite{covito_transient_2018}:
\begin{alignat}{2} \label{eq:semimetal_Gamma}
    \Gamma(E) &= \Gamma_0 &&\text{if } |E-E_{FM}| > W_L \\
    &= \Gamma_0 \frac{|E-E_{FM}|}{W_L} \quad &&\text{otherwise}.
\end{alignat}
First, we solve for the semimetal CNL relative to the wideband metal limit ($\text{CNL}_i$). Since MIGS is proportional to $\Gamma(E)$ (Eq. (4) in the main text \cite{MainPaper2026}), we may decompose the MIGS density $\rho_X(E)$ into the product of $\Gamma(E)$ and a coupling-independent multiplier,
\begin{equation}
\rho_X(E) \approx \rho_{X0}(E) \Gamma(E)    
\end{equation}
where $\rho_{X0}(E)$ is intrinsic to the semiconductor and independent of $\Gamma(E)$. The CNL definition from our main text \cite{MainPaper2026} can be expressed for the metal and semimetal as
\begin{equation}
    \begin{aligned} \label{eq:chargebalance_1}
    \int_{-\infty}^{\text{CNL}_i} \rho_{X0}(E)\Gamma_0 \d E &= \int_{\text{CNL}_i}^\infty \rho_{X0}(E) \Gamma_0 \d E \\
    \int_{-\infty}^{\text{CNL}} \rho_{X0}(E) \Gamma(E) \d E &= \int_{\text{CNL}}^\infty \rho_{X0}(E) \Gamma(E) \d E
\end{aligned}
\end{equation}
respectively. 
To simplify \cref{eq:chargebalance_1} analytically, we assume that $\rho_{X0}(E)$ is approximately constant in the bandgap, following Cowley and Sze \cite{cowley_surface_1965}. Defining the coupling deficit $\delta \Gamma(E) = \Gamma_0-\Gamma(E)$, \cref{eq:chargebalance_1} yields
\begin{equation} \label{eq:chargebalance}
    2\Gamma_0 (\text{CNL} -\text{CNL}_i) + \int_{\text{CNL}}^\infty \delta \Gamma(E) \d E - \int_{-\infty}^\text{CNL} \delta \Gamma(E) \d E  = 0
\end{equation}
which reduces to a quadratic equation 
solvable for CNL. The result is
\begin{equation} \label{eq:CNL_formula}
\text{CNL} -\text{CNL}_i = \begin{cases} \Delta + 2W_L\cdot \text{sgn}(\Delta) \left( \sqrt{1-\frac{|\Delta|}{2W_L}}-1 \right) & \text{if } |\Delta| \leq 3W_L/2 \\ \frac{W_L}{2} \cdot \text{sgn}(\Delta) & \text{if } |\Delta| > 3W_L/2 \end{cases}
\end{equation}
where $\Delta =E_{FM}-\text{CNL}_i$. Now that CNL has been established, the Fermi level $E_F$ of the system is derived from charge balance at the semimetal-TMD interface \cite{cowley_surface_1965}
\begin{equation}
    \frac{q^2d \rho_{X0}}{\epsilon} \int_{\text{CNL}}^{E_F}  \Gamma(E)\d E = E_{FM}-E_F
\end{equation}
where $d$ and $\epsilon$ are the thickness and dielectric constant of the semimetal-TMD gap. If $S = \left(1+\frac{q^2d\rho_{X0}\Gamma_0}{\epsilon} \right)^{-1}$ is the pinning factor for the normal metal, we find
\begin{equation} \label{eq:EF}
    E_F - \text{CNL} = \begin{cases} 
    S\delta  & \text{if } |\delta| > \frac{W_L}{1-S} \\
    \delta + \text{sgn}(\delta) \left(w - \sqrt{w^2 + 2W_L|\delta|-W_L^2} \right) & \text{if } W_L<|\delta| \leq \frac{W_L}{1-S} \\
    \delta + \text{sgn}(\delta) \left(w-\sqrt{\delta^2+w^2} \right) & \text{if } |\delta| \le W_L \end{cases}
\end{equation}
where $\delta = E_{FM}-\text{CNL}$ and $w = \frac{SW_L}{1-S}$.

Solutions to \cref{eq:CNL_formula,eq:EF} are plotted in \cref{fig:semimetal_2}, revealing the intricate physics of semimetal contacts. In a standard metal contact (red line), high MIGS density pins $E_F$ near the intrinsic $\text{CNL}_i$. However, a semimetal contact introduces two de-pinning mechanisms due to its linear dispersion near the Dirac or Weyl point: (1) there is no pinning at $E_{FM} = \text{CNL}_i$ because a semimetal contact has no MIGS at the Dirac or Weyl point (2) the semimetal CNL ``follows" the semimetal Fermi energy $E_{FM}$ (\cref{eq:CNL_formula}) because a shift in semimetal $E_{FM}$ changes the shape of MIGS (\cref{fig:semimetal}). In contrast, a shift in $E_{FM}$ of a normal wideband metal has no effect on MIGS since $\Gamma(E) = \text{const}$; thus, the charge neutrality level of the metal contact is fixed at $\text{CNL}_i$. Once $|E_{FM}-\text{CNL}_i|$ exceeds the semimetal's linear dispersion range, the CNL saturates at $|\text{CNL}-\text{CNL}_i|=W_L/2$ because $\Gamma(E)$ becomes flat. In this regime, the linear region of the semimetal is `inactive' because its energy range lies too far from $\text{CNL}_i$ to affect the band alignment (\cref{fig:semimetal_2}).

We may further simplify \cref{eq:CNL_formula,eq:EF} by deriving an \emph{effective pinning factor} $S' = \frac{\d E_F}{\d E_{FM}}$, averaged over the linear region $|E_{FM}-\text{CNL}_i|<3W_L/2$ of the semimetal. The result is
\begin{equation}
    S' \approx \frac{1+2S}{3}
\end{equation}
where $S$ is the pinning factor of a metal with similar coupling (\cref{fig:semimetal_2}). Hence, a semimetal causes the overall Fermi level $E_F$ to move toward $E_{FM}$ even for strongly pinned systems ($S\approx 0$), because the normal range of metal pinning factors $0\le S \le 1$ corresponds to an improved range of \emph{effective} semimetal pinning factors $\frac{1}{3} \le S' \le 1.$ Therefore, weak FLP is caused by the shift of CNL toward $E_{FM}$ (\cref{eq:CNL_formula}) rather than a true reduction in pinning strength. This CNL shift is illustrated in \hyperref[fig:semimetal]{Fig.~S3(b-d)}, where increasing the semimetal workfunction from 3.8 eV to 6.5 eV shifts the CNL from 0.27 eV (\hyperref[fig:semimetal]{Fig.~S3(b)}) to $-0.71$ eV (\hyperref[fig:semimetal]{Fig.~S3(d)}). The full numerical calculation of \cref{fig:semimetal} thus predicts $\frac{\d \text{CNL}}{\d E_{FM}}\approx \frac{0.27\text{ eV}+0.71\text{ eV}}{6.5\text{ eV}-3.8\text{ eV}}=0.36$, in excellent agreement with our analytical prediction of $\frac{\d \text{CNL}}{\d E_{FM}}\approx 1/3$ (\cref{eq:CNL_formula}).

In summary, we showed that semimetals effectively ``bypass" FLP by changing the ratio of conduction-to-valence band MIGS, rather than by simply decreasing their density. However, the efficacy of semimetal contacts can still be compromised by chalcogen vacancies, where DIGS dominate and pin $E_F$ to the chalcogen vacancy energy (\cref{sec:s2}). By deriving an \emph{effective} pinning factor for semimetals, our model identifies a fundamental advantage: semimetal-on-TMD contacts intrinsically outperform traditional metal-on-TMD contacts of similar workfunction and coupling strength. These findings establish semimetal contacts as a uniquely effective approach to improve strongly pinned metal-semiconductor contacts.

\bibliography{zotero_supp,manual_supp}